\documentclass[twocolumn,english,aps,prx,footnote,superscriptaddress,showpacs,lengthcheck,longbibliography]{revtex4-2}
\usepackage[T1]{fontenc}
\usepackage{lmodern}
\usepackage[latin9]{inputenc}
\usepackage{color}
\usepackage{babel}
\usepackage{array}
\usepackage{textcomp}
\usepackage{amsmath}
\usepackage{amssymb}
\usepackage{graphicx}
\usepackage[unicode=true,
 bookmarks=false,
 breaklinks=true,pdfborder={0 0 1},backref=false,colorlinks=false]
 {hyperref}
\usepackage{breakurl}

\makeatletter

\newcommand{\lyxmathsym}[1]{\ifmmode\begingroup\def\b@ld{bold}
  \text{\ifx\math@version\b@ld\bfseries\fi#1}\endgroup\else#1\fi}

\providecommand{\tabularnewline}{\\}

\usepackage{amsthm}
\usepackage{array}
\usepackage{float}
\usepackage[percent]{overpic}
\usepackage[nottoc, notlof, notlot]{tocbibind}
\usepackage{cases}
\usepackage{fullpage}
\usepackage{color}
\usepackage{subfigure}

\usepackage{dcolumn}
\usepackage{bm}

\makeatother

\begin{document}

\title{Quantum Key Distribution with a Hand-held Sender Unit}

\author{Gwenaelle Vest}
\affiliation{Ludwig-Maximilians-Universit\"at, 80799 M\"unchen, Germany}

\author{Peter Freiwang}
\affiliation{Ludwig-Maximilians-Universit\"at, 80799 M\"unchen, Germany}
\affiliation{Munich Center for Quantum Science and Technology (MCQST), 80799 M\"unchen,
Germany}

\author{Jannik Luhn}
\affiliation{Ludwig-Maximilians-Universit\"at, 80799 M\"unchen, Germany}

\author{Tobias Vogl}
\affiliation{Ludwig-Maximilians-Universit\"at, 80799 M\"unchen, Germany}
\affiliation{\inputencoding{latin1}University of Cambridge, CB3 0HE Cambridge,
United Kingdom}
\inputencoding{latin1}%
\affiliation{Friedrich-Schiller-Universit\"at Jena, 07745 Jena, Germany}
\selectlanguage{english}%

\author{\inputencoding{latin9}Markus Rau}
\inputencoding{latin9}%
\affiliation{Ludwig-Maximilians-Universit\"at, 80799 M\"unchen, Germany}

\author{Lukas Knips}
\email{lukas.knips@mpq.mpg.de}
\selectlanguage{english}%
\affiliation{Ludwig-Maximilians-Universit\"at, 80799 M\"unchen, Germany}
\affiliation{Munich Center for Quantum Science and Technology (MCQST), 80799 M\"unchen,
Germany}
\affiliation{Max-Planck-Institut f\"ur Quantenoptik, 85748 Garching, Germany}

\author{Wenjamin Rosenfeld}
\affiliation{Ludwig-Maximilians-Universit\"at, 80799 M\"unchen, Germany}
\affiliation{Munich Center for Quantum Science and Technology (MCQST), 80799 M\"unchen,
Germany}

\author{Harald Weinfurter}
\email{h.w@lmu.de}
\selectlanguage{english}%
\affiliation{Ludwig-Maximilians-Universit\"at, 80799 M\"unchen, Germany}
\affiliation{Munich Center for Quantum Science and Technology (MCQST), 80799 M\"unchen,
Germany}
\affiliation{Max-Planck-Institut f\"ur Quantenoptik, 85748 Garching, Germany}

\begin{abstract}
Quantum key distribution (QKD) is a crucial component
for truly secure communication, which enables
to analyze leakage of information due to eavesdropper attacks. While
impressive progress was made in the field of long-distance implementations,
user-oriented applications involving short-distance links have mostly
remained overlooked. Recent technological advances in integrated photonics
now enable developments towards QKD also for existing hand-held communication
platforms. In this work we report on the design and evaluation of a hand-held free-space QKD system
including a micro-optics based sender unit. This system implements
the BB84-protocol employing polarization-encoded faint laser pulses
at a rate of $100\,\mathrm{MHz}$. Unidirectional beam tracking and
live reference-frame alignment systems at the receiver side enable
a stable operation over tens of seconds when aiming the portable transmitter to the receiver input by hand from a distance of about half a meter. The user-friendliness of our
system was confirmed by successful key exchanges performed by different
untrained users with an average link efficiency of about $20\,\mathrm{\%}$
relative to the case of the transmitter being stationarily mounted
and aligned. In these tests we achieve an average quantum bit error
ratio (QBER) of $2.4\,\%$ and asymptotic secret key rates ranging
from $4.0\,\mathrm{kbps}$ to $15.3\,\mathrm{kbps}$. Given its compactness,
the versatile sender optics is also well suited for integration into
other free-space communication systems enabling QKD over any distance.
\end{abstract}
\maketitle

\section{Introduction}

Quantum Key Distribution (QKD) ~\cite{Bennett1984a,Tittel2002,Scarani2009a,Diamanti2016,pir2019advances,RevModPhys.92.025002}
enables the exchange of a secret cryptographic key between two authenticated
parties with the security based on fundamental physical laws.
Over the years QKD has evolved from an initial laboratory experiment~\cite{Bennett1992}
into a mature technology and fiber-based systems are already commercially
available. By extending the range of secure point-to-point fiber~\cite{Leverrier2013,Yin2016a,PhysRevLett.121.190502}
and free-space links~\cite{Schmitt-Manderbach2007a,Nauerth2013d}
an essential step towards large-scale quantum networks has been achieved.
More recently a satellite-to-ground key exchange has been demonstrated
on a global scale~\cite{Liao2017,PhysRevLett.120.030501}, and a
trusted-node based link was set up connecting Chinese cities over
a distance of $2000\,\mathrm{km}$ \cite{Zhang2018,pir2019advances}.

Yet, QKD has also a remarkable potential for secure short-distance communication tasks, e.g., for secure card-less payments~\cite{Duligall2006}, safe connected homes~\cite{Elmabrok:18} or user access to a quantum network~\cite{Dynes2013}. In these scenarios, a mobile device storing sensitive data could exchange keys with any static node within the network. The resulting keys could be consumed immediately to securely upload content or be earmarked for later use. Such applications, however, require user-friendly QKD sender units which are ideally already integrated into daily use mobile devices like mobile phones, tablets or laptops.

The developments in the field of integrated optics and microelectronics have opened up new possibilities for the miniaturization of QKD hardware. Such technological advances have led to the demonstration of miniaturized components~\cite{Li2014} and short-range free-space QKD optical links in static~\cite{Duligall2006,Benton2010,rarity_low_cost_sender} as well as in a hand-held~\cite{Chun2017} configuration. Based on highly integrated, on-chip QKD components~\cite{Sibson2015,Ma2016,Ziebell2015,Orieux2016,PhysRevX.8.021009_2018,PhysRevX.10.031030_2020,avesani2021daylight}, the capabilities for generating key material at an impressive rate in a full encryption scheme have been reported~\cite{Shields_2021_integrated_photonics}. Despite this great technological progress, all QKD hardware so far is only suited for stand-alone operation and still lacks integrability into portable communication devices.

In this work we combine these technological advances in miniaturizing all optical and electronical components for a hand-held QKD transmitter module. We detail its design and evaluate its performance in a realistic, fully hand-held free-space key exchange based on the BB84 protocol over a distance of about half a meter. Our micro-optics module generates attenuated laser pulses ($\lambda=850\,\mathrm{nm}$) with four different polarizations at a repetition rate of $100\,\mathrm{MHz}$. A visible beacon laser ($\lambda=680\,\mathrm{nm}$), overlapped with the QKD beam path, provides visual feedback during the aiming procedure and enables efficient unidirectional beam tracking by the receiver station as well as synchronization. In the current version, a smartphone placed on top of the transmitter communicates its orientation to establish a shared reference frame for polarization analysis of the detected QKD signals. The achieved user-friendliness and technological readiness of our system highlight the potential of the approach as well as its prospects for deployment in virtually all free-space scenarios.

\section{Miniature sender unit\label{sec:alice}}

\subsection{Design}

The present work focuses on integrating a QKD transmitter unit within
hand-held host systems such as smartphones. This requires a very compact
optical architecture with low power consumption implementing a simple
protocol with preferably only a small number of active components.
We thus chose the BB84 protocol with polarization encoding which can
be realized with standard laser sources and passive components only.

The four polarization states are generated by four laser diodes, respectively,
and in order to erase any information about the location of the source,
the output of the diodes is overlapped by a waveguide chip into one
spatial mode. This architecture achieves a high mechanical and thermal
stability and requires less than $2\times2\,\mathrm{mm^{2}}$ transversally
and $35\,\mathrm{mm}$ in longitudinal direction~\cite{Vest2014}.
The operating wavelength of $850\,\mathrm{nm}$ provides high transmission
through the atmosphere for free-space applications and a wide availability
of inexpensive, high-efficiency single-photon detectors operating
at non-cryogenic temperatures. 

\begin{figure*}
\subfigure{\label{fig:AliceBild}\includegraphics[width=0.75\textwidth]{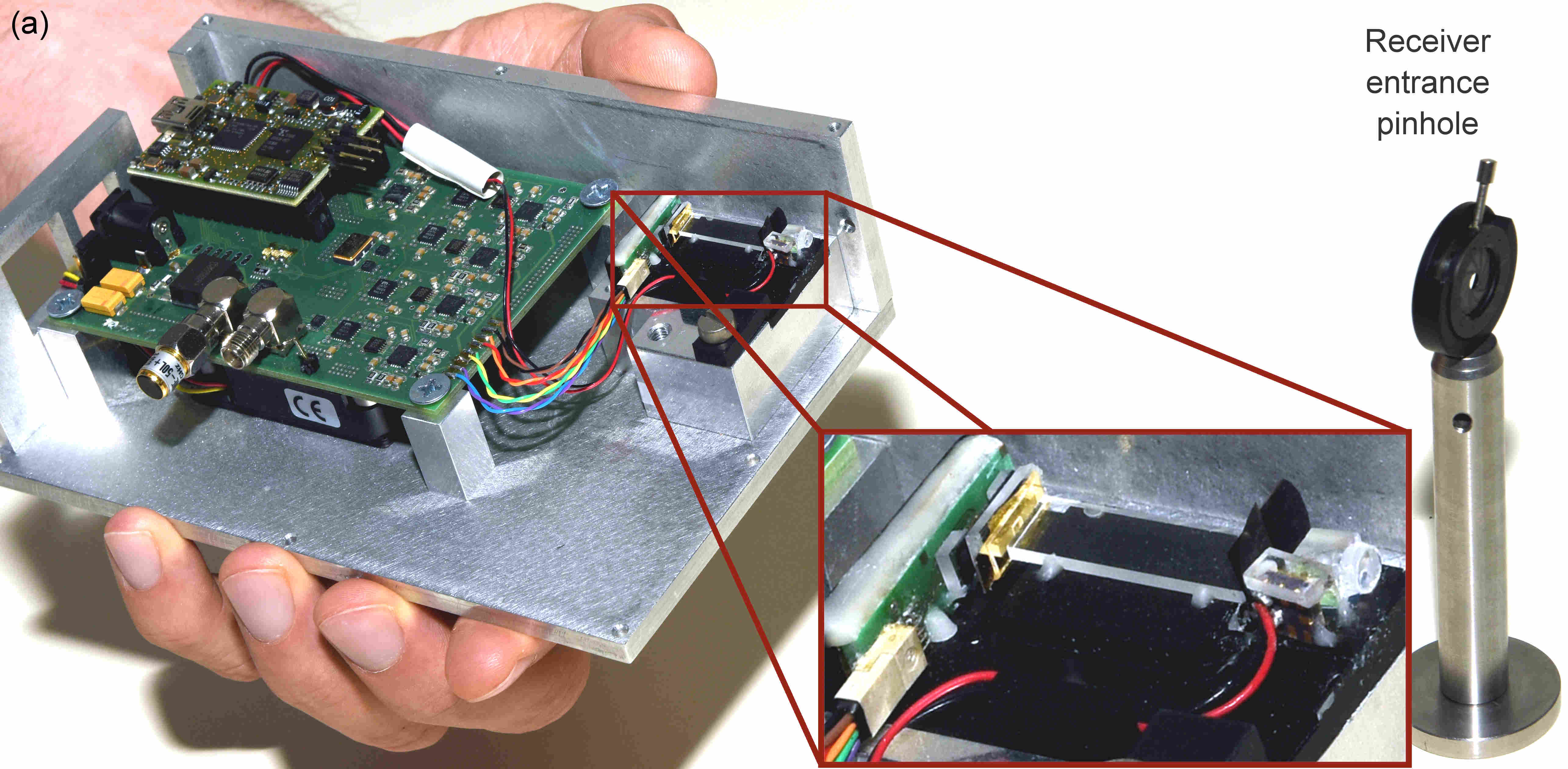}}
\begin{centering}
\vspace{0.4cm}
\par\end{centering}
\centering

\begin{minipage}[b][1\totalheight][t]{5.5cm}%
\subfigure{\label{fig:AliceLayout}\includegraphics[height=5cm]{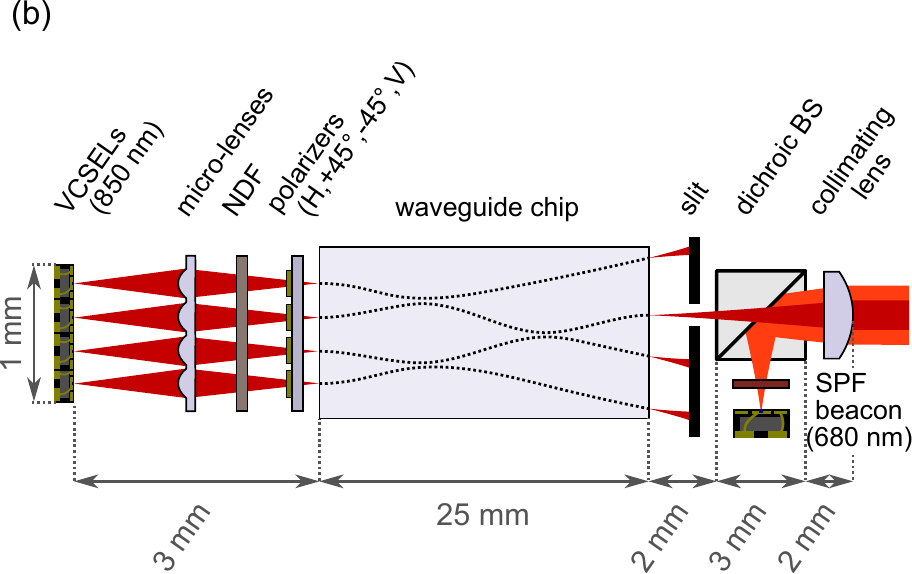}}%
\end{minipage}\hspace{4cm}%
\begin{minipage}[b][1\totalheight][t]{5.5cm}%
\subfigure{\includegraphics[height=5cm]{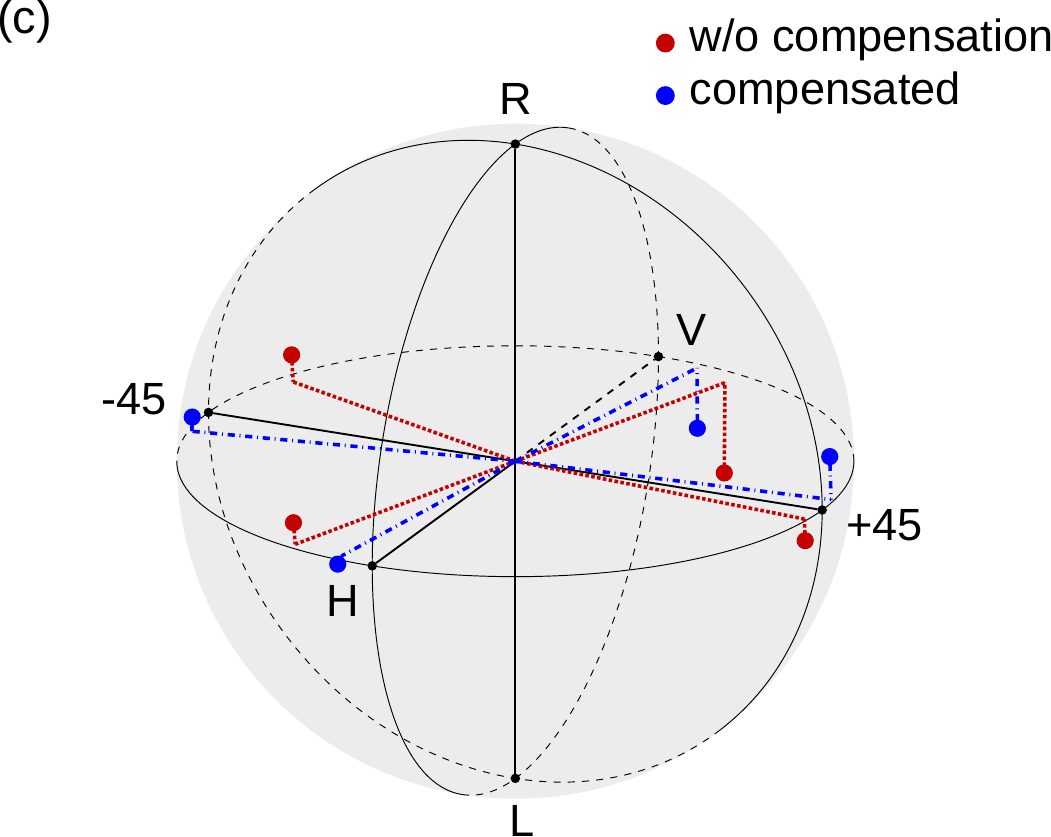}\label{fig:AliceStates}}%
\end{minipage}

\caption{\label{fig:alice-1}The hand-held transmitter unit. (a) Picture of
the mobile transmitter unit featuring a micro-optical bench (inset)
with a size of $30\times6.5\times5\,\mathrm{mm^{3}}$ (together with
PCB and connector $35\times20\times8\,\mathrm{mm^{3}}$) for generating
BB84 states and its dedicated electronics. The surrounding case has
a size of $15.5\times9\times5\,\mathrm{cm^{3}}$. (b) Schematic of
the optics architecture. Pulses centered around $850\,\mathrm{nm}$
and exhibiting a low degree of polarization are produced at $100\,\mathrm{MHz}$
rate by an array of four vertical-cavity surface-emitting lasers (VCSELs).
The pulses are attenuated by a neutral density filter (NDF), polarized
by an array of wire-grid polarizers and coupled into a low-birefringence
waveguide chip containing three directional couplers. A micro-lens
array enables here a coupling efficiency of about $20\,\%$. The desired
output of the waveguide is selected by a thin slit. An additional
beacon laser ($\lambda=680\,\mathrm{nm}$) is spectrally filtered
by a short-pass filter (SPF) to reduce background emission and overlapped
with the $850\,\mathrm{nm}$ signal using a miniature beamsplitter
to provide visual feedback during the aiming procedure and to allow
for synchronization as well as efficient beam tracking on the receiver's
side. (c) Polarization states as measured by quantum tomography at
the output of the transmitter unit (red) and after compensation by
the set of waveplates in the receiver (blue).}
\end{figure*}

The optical module (Fig.~\ref{fig:AliceBild}) was put together with
the electronics board into an aluminum case. The layout of the QKD
sender module is shown in Fig.~\ref{fig:AliceLayout} (for details
on components and assembly, see App.~\ref{app-sec:Components}).
As light sources we employ an array of four single-mode vertical-cavity
surface-emitting lasers (VCSELs) with a spacing of $250\,\mathrm{\lyxmathsym{\textmu}m}$.
The VCSELs are operated in the strong modulation regime, yielding
low background optical pulses with a duration of only $200\,\mathrm{ps}$
at $100\,\mathrm{MHz}$ repetition rate. Under strong and short carrier
injection, transient effects lead to a low degree of polarization
(DOP $<0.5$), allowing for generation of the four BB84 states ($H$,
$V$, $+45$, $-45$) by using polarization filters. For this purpose,
we fabricated four gold wire-grid polarizers~\cite{Bird1960,Guillaumee2009}
matching the $250\,\mathrm{\lyxmathsym{\textmu}m}$ pitch of the lasers
and exhibiting extinction ratios better than $1000:1$~\cite{Melen2015}.
The spatial overlap of the polarized beams is ensured by three directional
couplers integrated in a low-birefringence alumino-boro-silicate waveguide
chip manufactured via femtosecond laser writing~\cite{Davis1996,Gattass2008,Valle2009}.

Finally, the signal pulses are superimposed with a bright, visible
$680\,\mathrm{nm}$ beacon laser using a dichroic beamsplitter cube
and collimated with a small aspheric lens ($f=4.9\,\mathrm{mm}$).
The beacon light helps the user to aim at the receiver, and allows
for efficient beam tracking (see Sec.~\ref{sec:bob}). A modulation
of this beam ($100\,\mathrm{MHz}$) ensures an accurate synchronization
with the receiver.

The assembled optical module has a size of $35\times20\times8\,\mathrm{mm^{3}}$
where the large lateral extension is mainly determined by the footprint
of the electric connector and of the printed circuit board (PCB, $20\times6\,\mathrm{mm^{2}}$)
onto which the VCSEL array is mounted, as well as by the width of
the glass chip containing the multiple photonic circuits. The control
electronics, implemented on a single PCB ($96\times60\times18\,\mathrm{mm^{3}}$),
includes four laser drivers and pulse generators together with an
FPGA. The latter enables communication with a PC or a smartphone for
controlling the device parameters. For this proof-of-principle demonstration
the sequence of random bits determining the state to be sent was generated
by a pseudorandom number generator (Java SecureRandom class) and uploaded
to the FPGA. With a current storage capacity of $131056\times2$ bits,
the random pattern was then repeated periodically every $1.3\,\mathrm{ms}$.\textcolor{black}{{}
The physical size of the device can surely be reduced during industrialization
while the random numbers ideally have to be produced in real time
by a quantum random number generator.}

\subsection{Characterization of the sender unit\label{subsec:alice-Characterization}}

The polarization states emerging from the sender were characterized
by complete state tomography in order to evaluate their quality. Fig.~\ref{fig:AliceStates}
shows the results as red dots on the Poincar\'e sphere, which come close
to the desired set of states up to a rotation resulting from birefringence
in the waveguide chip (see App.~\ref{app-sec:Output-states} for
details). This rotation can be almost fully compensated by a set of
external half- and quarter-waveplates for all four states simultaneously.
A very good agreement between the simulation (not shown) and the experimentally
achieved compensation (Fig.~\ref{fig:AliceStates}, blue dots) was
observed. Alice's compensated states resulted in a source-related
quantum bit error ratio (QBER) $e_{\mathrm{source}}=1.5\,\mathrm{\%}$.
Furthermore, the tomographic measurements (see Appendix Tab.~\ref{tab:output-states}) also allow to determine how well the basis states are mutually conjugated using the so called preparation quality~\cite{Tomamichel2012}
\begin{equation}
\text{q}=-\text{\ensuremath{\log}}_{2}(\mathop{\mbox{max}}_{\{\psi_{H,V},\psi_{\pm45^{o}}\}}\left(\mid\langle\psi_{H,V}\mid\psi_{\pm45^{o}}\rangle\mid^{2}\right)).\label{eq:quality-factor}
\end{equation}
For our sender module, the maximum of the four products to be calculated is found for ${\mid\langle\psi_{V}\mid\psi_{+45^{o}}\rangle\mid^{2}}$ resulting in the reduced value ${\text{q} = 0.75}$ ($\text{q}_{\text{ideal}} = 1.0 $) mainly because of an angle misalignment in the production of the polarizers. Also imperfections due to the polarization-dependent loss (PDL) in the waveguide chip and other
optical components of the sender contribute to the preparation quality.

Finally, we have investigated the robustness of our implementation
to side-channel attacks~\cite{Nauerth2009,Rau2015}, which would
become possible due to residual distinguishability in any other degree
of freedom besides the polarization (see App.~\ref{app-sec:Components}).
Our high-speed electronics featuring delay lines with $5\,\mathrm{ps}$
resolution, together with the fast optical response of the VCSELs
allow us to achieve excellent temporal overlap of the pulses, while
the optimized design of the waveguide results in essentially perfect
spatial overlap. We observed a spectral distinguishability of the
4 off-the-shelf VCSELs due to fabrication imperfections, however,
for our proof-of-principle demonstration we accept this open side channel. When commercializing such systems in the future, this could be avoided,
e.g., by proper selection of the diodes prior to the assembly of the
module \cite{Nauerth2009}, by the use of MEMS-tunable structures~\cite{Davani2012a,Paul2014} or by spectral filtering once the spectra of the four diodes start overlapping.

\section{Free-space receiver with real-time alignment capabilities\label{sec:bob}}

Our free-space receiver (Fig.~\ref{fig:bob}) includes a standard
polarization analysis unit (PAU) capable of analyzing Alice's states
and featuring several extensions strongly simplifying the operation
during a hand-held key exchange. In particular, a dynamic alignment
system compensates for beam wander due to user's hand movements and
thereby ensures a stable optical link with the hand-held platform.
Furthermore, it provides an additional audio feedback to the user
allowing to maximize the transmission. Finally, an active reference
frame alignment compensates for varying rotations of the sender during
operation. 

\subsection{Polarization analysis}

\begin{figure*}
\centering{}\includegraphics[width=0.85\textwidth]{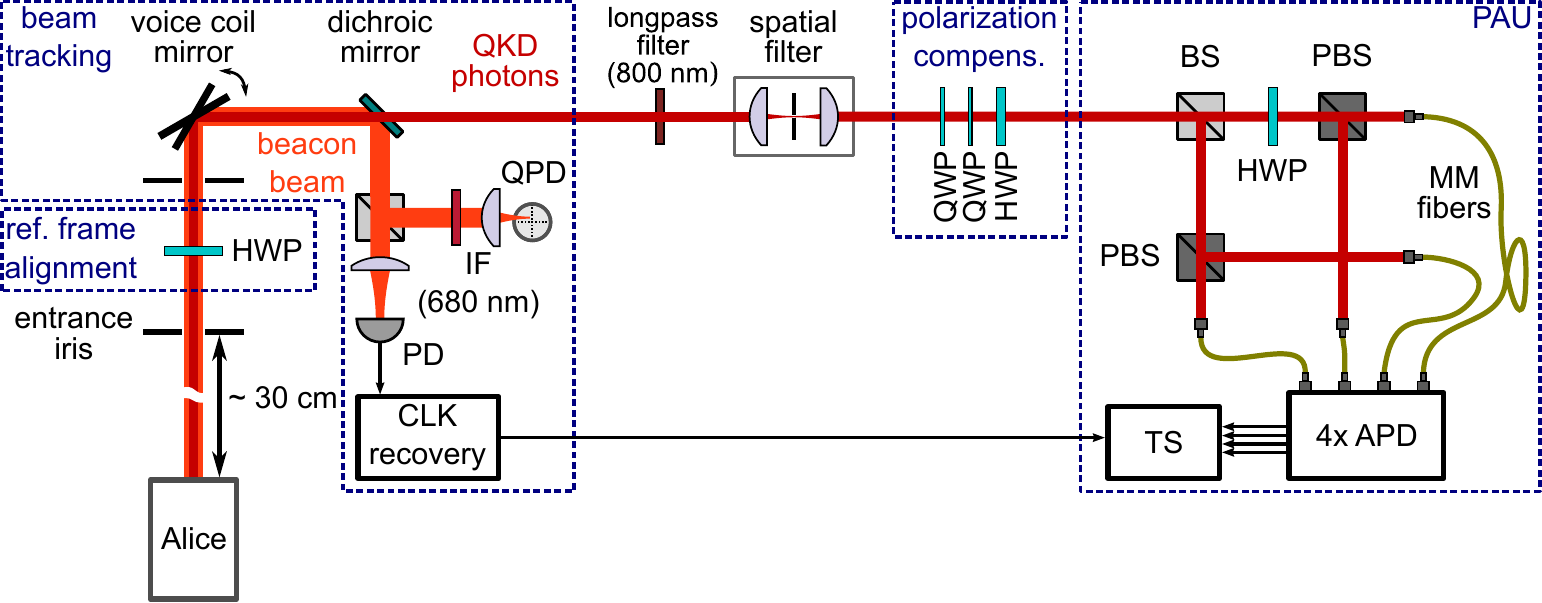}
\vspace{20pt}
 \caption{\label{fig:bob}Architecture of the free-space receiver. A polarization
analysis unit (PAU) discriminates between four BB84 states. Prior
to the spectral and spatial filtering of the attenuated pulses, a
motorized half-waveplate (HWP) rotates the reference frame of the
PAU depending on the sender unit's orientation. The latter is recorded
by a smartphone placed on top of the sender unit and transmitted over
WLAN. Polarization rotations occurring in the sender and receiver
are compensated by a set of motorized waveplates. Beam tracking is
performed on the beacon laser using a quadrant photodiode (QPD) and
an electrically driven voice coil mirror. The beacon is additionally
modulated at $50\,\mathrm{MHz}$ allowing for clock recovery with
a fast photodiode (PD). PD: photodiode; BS: beamsplitter; PBS: polarizing
beamsplitter; MM: multi-mode fiber; APD: avalanche photodiode; TS:
time-tagging (timestamp) unit.}
\end{figure*}

The PAU of the BB84 receiver device (see Fig.~\ref{fig:bob}) consists
of a 50:50 beamsplitter for the passive basis choice, two polarizing
beamsplitters and four actively quenched, fiber-coupled avalanche
photodiodes (APD, \textit{SPCM-AQ4C}, Perkin Elmer) with a specified
detection efficiency $\eta\approx38\,\mathrm{\%}$. Here, one has
to take into account the birefringent phases of the involved optical
components (gold and dichroic mirrors) leading to a rotation of the
incoming polarization states. This was analyzed separately by performing
a state tomography of well-defined input states directly at the entrance
of the PAU. With this and the knowledge of the polarization states sent by the transmitter (Sec.~\ref{subsec:alice-Characterization}) a unitary transformation which minimizes the detected QBER can be calculated and implemented by a set of motorized waveplates ($\lambda/4,\lambda/4,\lambda/2$). Full compatibility with any transmitter unit thus could be achieved
by, e.g., communicating the sender characteristics prior to the key
exchange.

\subsection{Reference-frame alignment}

A well-known challenge for moving platforms is that the transmitter
and the receiver in general do not share the same reference frame.
This problem can be overcome by using intrinsically rotation-invariant
states~\cite{D'Ambrosio2012a}, reference-frame-independent protocols
tolerating slow rotations~\cite{Zhang2014}, calibration of the channel
using strong classical light~\cite{Xavier2011}, or using information
extracted from the signal itself~\cite{Higgins2018}. Following our
quest towards a practical system under the conditions where only the
sender reference frame may tilt (stationary receiver), we use a motorized
half-waveplate in front of the receiver to perform a dynamic rotation
of Alice's states according to the current orientation of the sender
unit. The changes in the tilt of the user's hand are retrieved from
the attitude sensor of a smartphone placed on top of the transmitter.
Changes exceeding the specified resolution of the sensor ($1\,^{\circ}$)
are sent to the receiver over WLAN with a refresh rate of about $10\,\mathrm{Hz}$.

\subsection{Beam tracking}

It was recently shown that the pointing angle of the input beam with
respect to the optical axis of a free-space PAU opens a receiver side-channel
and thus is a critical security factor~\cite{Rau2015,Sajeed2015}.
The probability to detect each of the four BB84 states may indeed
vary with the angle of incidence, which could be exploited by an eavesdropper.
In order to avoid this potential side-channel we employed a spatial
filter. It restricts the possible incidence angles into the PAU to
maximally $\pm0.08\,^{\circ}$ and thereby guarantees equal coupling
efficiency into all four detectors. To enable hand-held operation
in spite of this significant restriction we implemented a beam steering
system which extends the acceptance range for the sender to $\pm3\,^{\circ}$.
For this purpose, the beacon beam is split off the quantum signals
using a dichroic mirror and its pointing angle is retrieved using
a quadrant photodiode. The error signal is fed to a voice coil mirror
featuring a bandwidth of about $800\,\mathrm{Hz}$.

\section{Quantum Key Distribution\label{sec:qkd}}

The performance of the whole system was first evaluated with the sender
unit being mounted about $30\,$cm away from the entrance iris of
the receiver. This allows the determination of the essential security
parameters, such as the mean photon number $\mu$ and the overall
transmission $T$ through the quantum channel and the receiver to
the detectors. In particular, for a given transmission there exists
an optimal $\mu$ maximizing the secure key rate. For the free-space
channel of $30\,$cm in air at our working wavelength we could safely
assume negligible absorption. Losses occur when coupling into the
receiver due to absorption by optical elements and spectral filters,
as well as coupling through the spatial mode filter and into the fibers
of the detectors. The resulting transmission was measured for the
case of optimal alignment and amounted to $T_{\mathrm{Bob}}=40.9\,\%$
(App.~\ref{app:methods}). In the hand-held case the transmission
was generally lower, we define the relative hand-held efficiency $\xi\in[0,1]$
such that $T(t)=\xi(t)\cdot T_{\mathrm{Bob}}$.

To initiate the key exchange, a PC uploads the parameters for the
laser diodes as well as the random bit sequence ($131\mathrm{k}\times2\:\mathrm{bits})$
to the FPGA which then starts the sequence of attenuated pulses. Detection
events are recorded by a time-tagging unit of Bob's PAU and correlated
to the bit sequence sent by Alice to distill a raw key.

\subsection{Static configuration\label{subsec:Static-configuration}}

With the sender module mounted and aligned, we observed for a mean
photon number $\mu=0.042$ \footnote{For better comparability, in the static configuration we set $\mu$
to a value which was optimized and used for the hand-held case.}, a raw key rate (overall detection rate before sifting, $1.5\,\mathrm{ns}$
time window per pulse applied) of $R_{\mathrm{raw}}=649.5\,\mathrm{kbps}$
and a QBER of $e=2.1\,\%$. The QBER mainly results from imperfect
preparation of polarization states and a background contribution by
emission of the VCSELs below threshold. The contribution of the beacon
laser and detector dark counts was as small as $0.075\,\%$. With
the high transmission observed here, the simple GLLP~\cite{Gottesman2004}
scheme is still very efficient. To evaluate the final extractable secure key rate $R_{sec}$ we also account for imperfect preparation of the quantum states using the preparation quality q (Eq.~\ref{eq:quality-factor}), resulting in
\begin{align}
R_{\mathrm{sec}} & =R_{\mathrm{sift}}\Bigg[(1-\Delta)\left(\text{q}-H_{2}\left(\dfrac{e}{1-\Delta}\right)\right)\label{eq:rate-GLLP}\\
 & -f(e)\cdot H_{2}(e)\Bigg],\nonumber
\end{align}
where $R_{\mathrm{sift}}$ is the key rate after the basis reconciliation
(sifting), $\Delta=\frac{1-(1+\mu)e^{-\mu}}{T\cdot\eta\cdot(1-e^{-\mu})}$
the fraction of ``tagged'' pulses, $\eta$ the detector efficiency,
$H_{2}$ the binary entropy, and $T$ is the overall transmission
(in the static case $T=T_{\mathrm{Bob}}$). Assuming an error correction
efficiency $f(e)=1.22$, we obtained $R_{\mathrm{sec}}=103.2\,\mathrm{kbps}$
(without considering finite key effects).

\subsection{Hand-held operation}

The hand-held QKD tests were performed by four untrained users with
two attempts each. The user removed the sender unit from its pedestal
and aimed at the receiver entrance iris with the help of the visible
beacon laser. Acoustic feedback informed the user about the quality
of the aiming, where a low (high) pitch sound corresponded to a small
(big) deviation from optimal pointing, respectively. Evidently, it is possible for a person to aim well at a point. Thus, no additional pointing hardware is required in the sender. The small fluctuations due to unsteady hands are compensated by the receiver tracking system. 

In order to take the fluctuating transmission into account, additional
post-processing steps had to be included in the data analysis~\cite{Semenov2009,Semenov2010,Erven2012,Vallone2015}.
The method used here is based on defining a certain transmission threshold
value $T_{\mathrm{thr}}$ (corresponding to a link efficiency threshold
$\xi_{\mathrm{thr}}$) and discarding detection events where the transmission
is too low. For the accepted events we conservatively assume $T_{\mathrm{thr}}$
as the transmission value for the evaluation of $R_{\mathrm{sec}}$
via Eq.~(\ref{eq:rate-GLLP}). For each time bin $k$ (of $10\,\mathrm{ms}$)
we then obtain a raw key rate
\[
R_{\mathrm{raw}}(k,\xi_{\mathrm{thr}})=\begin{cases}
R_{\mathrm{raw}}(k) & \textrm{for}\:\xi(k)\geq\xi_{\mathrm{thr}}\\
0 & \textrm{for}\:\xi(k)<\xi_{\mathrm{thr}}
\end{cases}
\]
There is an optimal $T_{\mathrm{thr}}$ since a low threshold level
requires a high shrinkage of the key during privacy amplification,
also due to a typically higher QBER, while a high threshold leads
to discarding a large amount of data and thus a lower raw key rate.
As it turned out, this value did not vary much for the different trials
and could be defined in advance for a practical application.

\begin{figure}
\begin{centering}
\includegraphics[width=1\columnwidth]{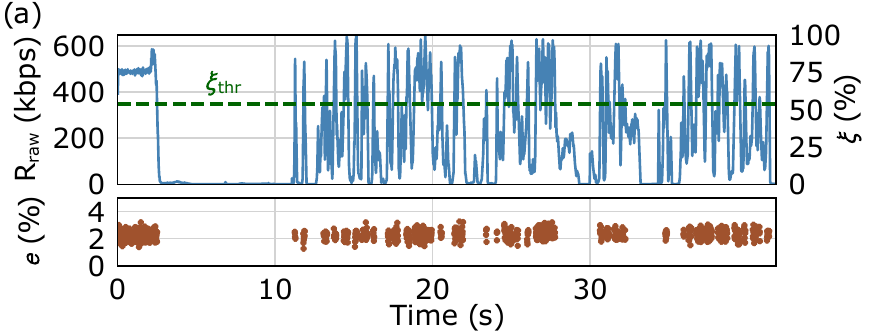} 
\par\end{centering}
\begin{centering}
\includegraphics[width=1\columnwidth]{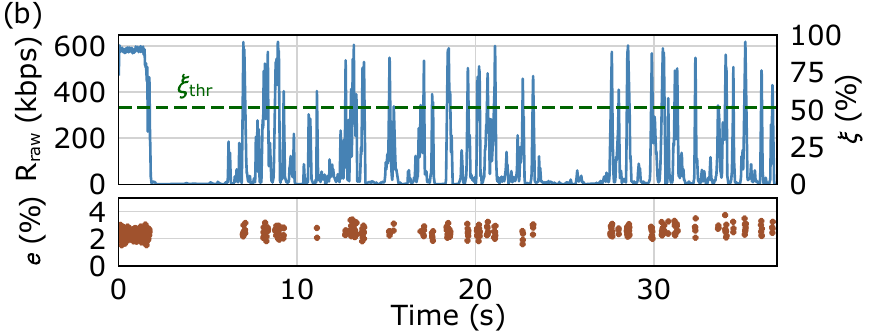}
\par\end{centering}
\centering{}\caption{\label{fig:trace-handheld01}Raw key rate $R_{\mathrm{raw}}$, relative
hand-held efficiency $\xi$ and QBER $e$ obtained in realistic key
exchanges for two different untrained users ((a), (b)) over a distance
of about $30\,\mathrm{cm}$. The rates are calculated for time bins
of $10\,\mathrm{ms}$. The threshold level $\xi_{\mathrm{thr}}$ is
represented by a dashed green line. }
\end{figure}

Figure~\ref{fig:trace-handheld01} shows traces of the raw key rate
together with its corresponding QBER for two trials of different users.
In order to also evaluate the time required to establish the connection
with the receiver, we started to record detection events while the
transmitter was still attached to the mount (but not necessarily optimally
aligned as in Sec.~\ref{subsec:Static-configuration}). The drop
in transmission corresponds to the moment where the user picked up
the sender, and the following period of low rate is due to the initial
pointing attempt. On average the initial pointing took $8.5$ secon\textcolor{black}{ds.
We set the average photon number per pulse to $\mu=0.042$, a value
supporting a good key generation performance despite the different
handling capabilities (shaking) of the users. }The results are presented
in App.~\ref{app:handheld-qkd-data} and show secret key rates between
$4.0\,\mathrm{kbps}$ and $15.3\,\mathrm{kbps}$, with an average
value of $7.1\,\mathrm{kbps}$ and a mean QBER of $2.4\,\mathrm{\%}$.
We evaluated for each trial the link efficiency $\xi_{\mathrm{link}}=\frac{1}{k_{\mathrm{link}}}\sum_{k}\xi(k)$
for the time intervals between first successful pointing and until
the respective trial was stopped (corresponding to $k_{\mathrm{link}}$
time bins). Averaging over all trials we obtain a value as high as
$\overline{\xi}_{\mathrm{link}}=21.1\,\mathrm{\%}$ confirming the
capabilities of our beam steering system.

The secret key rates could be further increased in the future by implementing
a decoy protocol~\cite{Lo2005}, an extension providing full security
against photon-number splitting attacks even for a higher mean photon
number per pulse. While our current electronics does not support this
functionality, we were still able to conduct an analysis by extracting
the relevant parameters from an additional experimental hand-held
dataset. This dataset of $7$ key exchanges with $3$ different users
was obtained for a significantly higher $\mu=0.153$ which would be
allowed by the decoy protocol. There we measured an average QBER of
$1.6\,\mathrm{\%}$, close to the source intrinsic error of the transmitter
unit of $1.5\,\mathrm{\%}$, thus confirming the quality of our automatic
reference-frame alignment. To estimate the possible performance, we
use the formalism introduced in Ref.~\cite{Ma2005} for a protocol
employing two decoy states (vacuum + weak decoy state). Assuming a
fraction of $97\,\mathrm{\%}$ for signal pulses and a mean photon
number for the weak decoy state of $\nu=0.077$ would allow for an
increased secret key rate with an average value exceeding $100\,\mathrm{kbps}$.
For a practical application it is also important that the key extraction
can be performed with a mobile platform within a reasonable time.
Preliminary tests with a smartphone showed sufficient processing capabilities
even for on-the-fly data processing for the range of detection rates
and QBER observed in this work.

\section{Conclusion and outlook\label{sec:Conclusion-Outlook}}

We developed a micro-optics based QKD transmitter unit and assembled
a prototype which features a very small footprint. The architecture
is optimized for BB84-like protocols with weak polarized laser pulses
at a wavelength of $850\,\mathrm{nm}$ and operates at $100\,\mathrm{MHz}$
repetition rate. The QKD beam is overlapped with a visible beacon
laser facilitating both the beam tracking on the receiver's side,
as well as clock synchronization between the two parties. Several
untrained users tested our integrated hand-held transmitter unit and
obtained on average a key generation rate of $7.1\,\mathrm{kbps}$.

Further measurements showed that the secret key rate can be increased
by more than one order of magnitude. Equipping the sender electronics with such decoy capabilities should lead to only an insignificant increase of form factor and power consumption.

The ability to generate sub-nanosecond pulses, assuming that the electronics and single-photon detectors can also be upgraded to such speeds, suggests
that the optoelectronics platform is suitable even for gigahertz operation.
This will put the key rates into the Mbit/s range thereby enabling
to exchange a significant amount of key during a short time.

Owing to its compact architecture, the presented transmitter unit
could easily be integrated in other existing optical communication
systems. It could provide secure key exchange for a variety of applications
like free-space links in urban areas, drones and high-attitude platforms
or even small low-earth orbit satellites enabling global secure communication.
\begin{acknowledgments}
The authors thank G. Corrielli, A. Crespi and R. Osellame at the Politecnico
di Milano for the design and fabrication of the waveguide array and
VI-Systems GmbH for providing the single-mode VCSELs. This project
was funded by the excellence cluster Nano-Initiative Munich (NIM),
by the EU project OpenQKD, by Deutsche Forschungsgemeinschaft (DFG,
German Research Foundation) under Germany\textquoteright s Excellence
Strategy \textendash{} EXC-2111 \textendash{} 390814868 and by the BMBF
Projects QUBE and QUBE-II.
\end{acknowledgments}

\appendix

\section{Optical transmitter unit}

\subsection{Components and architecture\label{app-sec:Components}}

\subsubsection*{VCSELs}

\begin{figure}
\begin{centering}
\includegraphics[width=1\columnwidth]{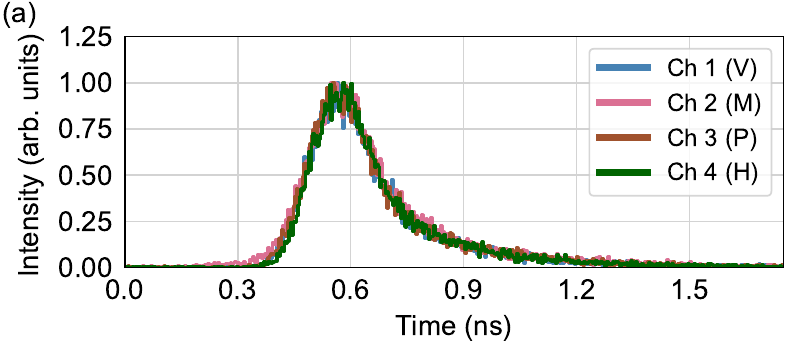}
\par\end{centering}
\begin{centering}
\includegraphics[width=1\columnwidth]{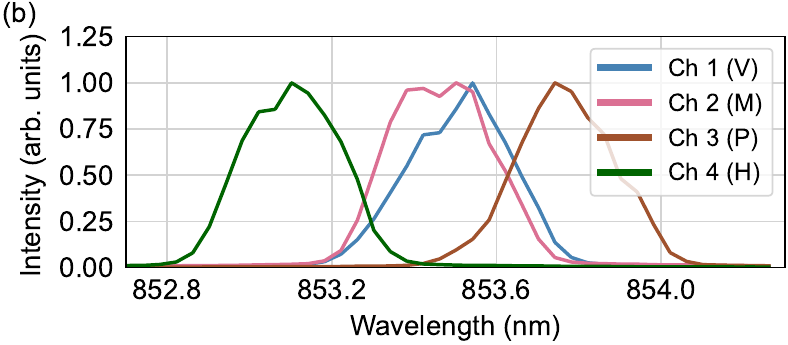}
\par\end{centering}
\caption{\label{fig:pulse-shapes-and-spectrum}Temporal and spectral properties
of Alice's output pulses. (a) Time-difference histogram of the photons
with respect to the $100\,\mathrm{MHz}$ clock showing the temporal
shape of the optical pulses generated by the four channels. Precise
synchronization of the channels and individual tailoring of the pulse
parameters result in a high temporal overlap. (b) Spectrum obtained
with a grating-based spectrometer with single-photon sensitivity featuring
a resolution of $125\,\mathrm{pm}$. Evidently, to avoid the spectral
side-channel, selected or tunable diodes have to be used for secure
applications.}
\end{figure}

The VCSEL-array used in the prototype was fabricated by VI-Systems
GmbH as a single chip ($1\times0.25\,\mathrm{mm^{2}}$) and features
four VCSELs with a low-threshold ($I_{b}=0.4\,\mathrm{mA}$) as well
as transverse and longitudinal single-mode emission around $850\,\mathrm{nm}$.
Each laser is operated in the strong modulation regime, \textit{\emph{i.e.}},
the diodes are biased below threshold and switched on for a short
time period by an intense RF pulse. These conditions ensure the generation
of bright, phase-randomized $200\,\mathrm{ps}$ long pulses with low
background emission in the off-state. The temporal overlap of the
pulses originating from different channels was retrieved from a time-difference
histogram (see Fig.~\ref{fig:pulse-shapes-and-spectrum}(a)), where
the arrival time of the attenuated pulse was measured with an avalanche
photodiode featuring a jitter of $30\,\mathrm{ps}$ and a time-to-digital
converter with a resolution of about $80\,\mathrm{ps}$. The spectrum
of these pulses was taken with a grating-based spectrometer with a
resolution of $125\:\mathrm{pm}$ (Fig.~\ref{fig:pulse-shapes-and-spectrum}(b)).
The observed spectral mismatch of the different laser diodes is less
than $0.7\,\mathrm{nm}$, but has to be further reduced by using selected
or tunable diodes. Clearly, for hardware like our sender module the open spectral side channel has to be closed but was accepted for this proof of principle experiment.

\subsubsection*{Nano wire-grid polarizers}

The geometry of the subwavelength nano-wire gratings was carefully
optimized using finite-difference time-domain simulations~\cite{Melen2015}.
A gold film deposited onto a thin glass substrate was milled by a
focused gallium ion beam. The choice of this manufacturing technique
is motivated by its high etching resolution, crucial for the observation
of extinction ratios above $30\,\mathrm{dB}$ (see Table~\ref{tab:polarizer-ER}),
and by its capability to precisely control the relative orientation
of the polarizers. Small polarization effects occurring within the
waveguide chip could be determined beforehand and thus be pre-compensated
by fabricating polarizers with certain linear polarizations, in order
to obtain output states as close as possible to the states required
for the BB84 protocol. The wire-grid polarizers feature a transmission
of about $9\,\%$ for the selected polarization, while the orthogonal
polarization component is strongly reflected. As the VCSELs are sensitive
to optical feedback, we placed a thin neutral density filter with
a transmission of ca. $8\,\%$ between the VCSEL array and the polarizers
in order to avoid polarized retro-injection.

\begin{table}[h]
\centering{} %
\begin{tabular}{|c|c|c|c|c|}
\hline 
 & $H$ & $+45$  & $-45$  & $V$\tabularnewline
\hline 
ER  & 1800:1  & 1620:1  & 1200:1  & 1150:1 \tabularnewline
\hline 
\end{tabular} \caption{\label{tab:polarizer-ER}Extinction ratios (ERs) of the employed wire-grid
polarizers at $850\,\mathrm{nm}$, differences are due to variation
of the grid parameters~\cite{Melen2015}.}
\end{table}

\subsubsection*{Waveguide chip\label{app-subsec:Waveguide}}

\begin{figure}
\includegraphics{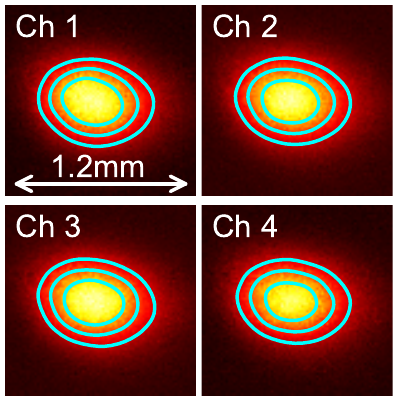}\caption{\label{fig:spatialMode}Spatial mode distribution of the four output
states of the sender unit. The pictures were taken with an EMCCD camera
at  a distance of about $50\,\mathrm{cm}$ from the sender's output
aperture. The cyan contour lines of relative optical intensity are
shown for facilitating visual comparison of the distributions.}
\end{figure}

The three directional waveguide couplers ensuring the spatial mode
overlap (see Fig.~\ref{fig:spatialMode}) of the four polarized beams
were implemented in a alumino-boro-silicate chip by femtosecond laser
writing~\cite{Davis1996,Gattass2008,Valle2009} by the Politecnico
di Milano \cite{Proc_Melen2016}. The low intrinsic birefringence
($\Delta n=7\cdot10^{-5}$) of the waveguides introduces a global
shift as small as $3\pi$ over the length of the chip. The polarization
dependence of the coupling ratios was minimized by engineering a 3D-layout~\cite{Heilmann2014}
of the couplers~\cite{Sansoni2012} resulting in only small changes
in the relative angles between the input polarization states after
propagation through the waveguide. See~\ref{app-sec:Output-states}
for characterization of the resulting output states.

\subsubsection*{Collimating optics}

For overlapping the signal and the beacon light a $3.5\times3.5\times3\,\mathrm{mm^{3}}$
dichroic beamsplitter from a commercial DVD-player was used. The transmission
for the $850\,\mathrm{nm}$ signal exceeds $99.8\,\%$, while the
reflection amounts to $50\,\%$ for the $680\,\mathrm{nm}$ beacon
beam. The aspheric lens collimating both beams has a diameter of $3\,\mathrm{mm}$
and a focal length of $4.9\,\mathrm{mm}$. This optics is well-suited
for short-distance operation and maintains a compact architecture.
Key exchanges over longer distances could be performed by combining
the miniature sender unit with a telescope or with optical terminals
of other free-space communication systems.

\subsubsection*{Micro-optics assembly}

The careful characterization of the micro-optics components was followed
by their precise positioning on the anodized aluminum micro-optical
bench using a custom-made vacuum gripper mounted onto a standard 6-axis
stage. Repeated stacking and gluing of the components using a UV-curing
adhesive ensured a good mechanical stability of the optical setup.
The distance between the optical parts was precisely controlled by
an optimal combination of silica spacers whose thicknesses were selected
according to Zemax simulations to achieve a high coupling efficiency.
After the assembly we observed a rather uniform coupling efficiency
close to $20\,\%$ across the VCSEL array. The current prototype has
a size of $35\times20\times8\,\mathrm{mm^{3}}$, although a size of
$30\times3\times3\,\mathrm{mm^{3}}$ seems feasible by exploiting
more suitable assembly techniques, a tighter cutting of the waveguide
glas substrate and especially by reducing the size of the connector
and of the PCB down to the minimal required area. 

\subsubsection*{Compatibility with high volume manufacturing}

Most of the elements (VCSELs, micro-lenses, spacers, filter, beamsplitter
and lens) are commercially available, and the few custom-made elements
require only standard fabrication processes and could in principle
be also manufactured in large series. For instance, the wire-grid
polarizers have been fabricated using FIB milling, while other groups
have demonstrated successful manufacturing of these elements using
nanoimprint~\cite{Ahn2005} or UV-lithography~\cite{Verrier2015,Liu2006}.
In the case of femtosecond laser writing, the scanning speed can be
as large as a few centimeters per second, and is thus viable for high
volume fabrication.

\subsection{Analysis of the polarization states\label{app-sec:Output-states}}

The Stokes components of the polarization states were reconstructed
by quantum state tomography entailing successive measurements in three
different bases $\left\{ H,V\right\} $,$\left\{ \pm45^{o}\right\} $
and $\left\{ R,L\right\} $:
\begin{align*}
S_{1} & =\frac{I_{H}-I_{V}}{I_{H}+I_{V}},\\
S_{2} & =\frac{I_{+45^{o}}-I_{-45^{o}}}{I_{+45^{o}}+I_{-45^{o}}},\\
S_{3} & =\frac{I_{R}-I_{L}}{I_{R}+I_{L}}.
\end{align*}
Due to the angle misalignment during the production of the polarizers, the sender output states (Tab.~\ref{tab:output-states}~a) are on the one hand jointly rotated and on the other hand also not perfectly mutually conjugated. While the former can be compensated at the receiver, the latter is accounted for during privacy amplification (Eq.~\ref{eq:rate-GLLP}) by the preparation quality q (Eq.~\ref{eq:quality-factor}). Tab.~\ref{tab:output-states} b) and c) detail the states detected by the receiver before and after compensation of possibly rotated states of the sender and further rotations due to birefringent components in the receiver using a set of waveplates before the PAU. This simplifies the sender significantly, while the receiver can easily adopt to different transmitter polarization state orientations by communicating the respective tomographic data at the begin of a key exchange.

\begin{table}[h]
\begin{centering}
\begin{tabular}{|c|c|c|c|c|}
\multicolumn{1}{c}{a)} & \multicolumn{1}{c}{} & \multicolumn{1}{c}{} & \multicolumn{1}{c}{} & \multicolumn{1}{c}{}\tabularnewline
\hline 
 & H & V & $+45^{o}$ & $-45^{o}$\tabularnewline
\hline 
 $S_{1}$  & 0.944 & -0.868 & 0.197 & -0.326\tabularnewline
\hline 
$S_{2}$ & -0.300 & 0.367 & 0.969 & -0.918\tabularnewline
\hline 
$S_{3}$ & 0.120 & -0.292 & 0.011 & 0.162\tabularnewline
\hline 
\end{tabular}
\par\end{centering}
\begin{centering}
\medskip{}
\begin{tabular}{|c|c|c|c|c|}
\multicolumn{1}{c}{b)} & \multicolumn{1}{c}{} & \multicolumn{1}{c}{} & \multicolumn{1}{c}{} & \multicolumn{1}{c}{}\tabularnewline
\hline 
 & H & V & $+45^{o}$ & $-45^{o}$\tabularnewline
\hline 
 $S_{1}$  & 0.938 & -0.855 & 0.102 & -0.234\tabularnewline
\hline 
$S_{2}$ & -0.134 & 0.094 & 0.926 & -0.858\tabularnewline
\hline 
$S_{3}^{*}$ & 0.319 & -0.509 & -0.362 & 0.457\tabularnewline
\hline 
\end{tabular}
\par\end{centering}
\begin{centering}
\medskip{}
\begin{tabular}{|c|c|c|c|c|}
\multicolumn{1}{c}{c)} & \multicolumn{1}{c}{} & \multicolumn{1}{c}{} & \multicolumn{1}{c}{} & \multicolumn{1}{c}{}\tabularnewline
\hline 
 & H & V & $+45^{o}$ & $-45^{o}$\tabularnewline
\hline 
 $S_{1}$  & 0.949 & -0.971 & -0.091 & -0.007\tabularnewline
\hline 
$S_{2}$ & 0.004 & 0.068 & 0.982 & -0.990\tabularnewline
\hline 
$S_{3}^{*}$ & 0.314 & 0.228 & 0.163 & 0.137\tabularnewline
\hline 
\end{tabular}
\par\end{centering}
\centering{}\caption{\label{tab:output-states} Tomographic data. a) Complete tomography
of the sender's output states measured with an additional arrangement
of a polarizer, QWP and an APD. The average degree of polarization
(DOP) is $0.990$ and the preparation quality (Eq.~(\ref{eq:quality-factor})) $\text{q} = 0.75$. b) Partial tomography of the sender's states as measured by the receiver without compensation. The modulus of the third Stokes parameter $S_{3}^{*}$ is calculated from $S_{1},S_{2}$ with the assumption of unity DOP, reasonably confirmed by the complete tomography of the sender (a) beforehand. The sign is inferred from the polarization transformation within the receiver known from additional measurements. c) Partial tomography of the sender's states as measured by the receiver with optimal compensation.}
\end{table}

\section{Methods\label{app:methods}}

\subsection{Measurement of the transmission}

The transmission through the receiver is defined by $T_{\mathrm{Bob}}=\frac{I_{\mathrm{PAU\text{*}}}}{I_{\mathrm{PAU}}}$
determined by taking the ratio of two count rates where $I_{\mathrm{PAU\text{*}}}$
is the average count rate of the light output by the Alice module
(static configuration, pulsed mode) detected by the four APDs of the
PAU, while $I_{\mathrm{PAU}}$ is obtained by coupling the light from
the output of the Alice module into a multimode fiber (coupling efficiency
of $83.3\,\%$) connected to one of the PAU APDs. These measurements
were performed for each of the four output polarizations and then
averaged resulting in $T_{\mathrm{Bob}}=40.9\,\%$.

\subsection{Determination of the mean photon number}

In order to determine the mean photon number $\mu$ we directly measured
the count rate at the receiver. From this number (at a given repetition
rate) $\mu$ can be extracted by using the receiver transmission,
the specified efficiency of the detectors ($\eta=38\,\mathrm{\%}$)
and accounting for the slightly polarization-dependent transmission and reflection
coefficients, respectively, of the BS in the PAU as well as the corrections of non-linearity
of the detectors at high count rates.

\section{Hand-held key exchange data\label{app:handheld-qkd-data}}

\begin{table}[!h]
\centering{}%
\begin{tabular}{|c||c|c|c|c|c|c|c|}
\hline 
User & Time  & Aiming & $\xi_{\mathrm{link}}$  &  $\xi_{\mathrm{thr}}$ & QBER & $R_{\mathrm{raw}}^{*}$  & $R_{\mathrm{sec}}$ \tabularnewline
 & s & s & \% & \% & \% & kbps & kbps\tabularnewline
\hline 
\hline 
1  & 30.5 & 8.0 & 34.5  & 53.8  & 2.3  & 140.3  & 15.3 \tabularnewline
2  & 31.0 & 4.0 & 13.9  & 51.2  & 2.6  & 43.9  & 4.0 \tabularnewline
3  & 33.0 & 17.0 & 20.6  & 51.2  & 2.2  & 76.1  & 8.4 \tabularnewline
4  & 40.5 & 6.0 & 19.0  & 45.2  & 2.6  & 69.3  & 5.3 \tabularnewline
1  & 33.5 & 9.5 & 16.9  & 60.7  & 2.4  & 46.3  & 5.4 \tabularnewline
2  & 41.0 & 6.0 & 20.5  & 62.0  & 2.3  & 41.3  & 5.0 \tabularnewline
3  & 61.0 & 10.0 & 20.2  & 52.9  & 2.3  & 68.6  & 7.2 \tabularnewline
4  & 39.5 & 7.5 & 23.3  & 61.9  & 2.5  & 59.6  & 6.4 \tabularnewline
\hline 
\multicolumn{1}{|c||}{Avg.} & 38.8 & 8.5 & 21.1 & 54.9 & 2.4 & 68.2 & 7.1\tabularnewline
\hline 
\end{tabular}\caption{\label{tab:handheld-summary01}Hand-held measurements obtained for
different users at a mean photon number $\mu=0.042$. $\xi_{\mathrm{link}}$
represents the link efficiency, i.e., the ratio between the average
rates observed in the hand-held case during link time and the rate
in the static configuration. $\xi_{\mathrm{thr}}$ is the link efficiency
threshold used for accepting data, $R_{\mathrm{raw}}^{*}$ the raw
key rate after the time and threshold filtering and $R_{\mathrm{sec}}$
the calculated asymptotic secure key rate according to Eq.~(\ref{eq:rate-GLLP}).}
\end{table}

\end{document}